\documentstyle[epsf]{elsart}

\begin{document}
\begin{frontmatter}
\title{ Vlasov calculations of nuclear ground states  }
\author[LNS]{Aldo Bonasera} \ead{bonasera@lns.infn.it} 
\author[JAERI]{Toshiki Maruyama} \ead{maru@hadron02.tokai.jaeri.go.jp}
\author[CT,LNS]{Massimo Papa} \ead{papa@lns.infn.it}
\author[JAERI]{Satoshi Chiba} \ead{chiba@hadron31.tokai.jaeri.go.jp}

\address[LNS]{
Istituto Nazionale Fisica Nucleare-Laboratorio Nazionale
del Sud, Via Santa Sofia 44, Catania 95123, Italy
}
\address[JAERI]{
Advanced Science Research Center, Japan Atomic Energy Research Institute,
Tokai, Ibaraki 319-1195, Japan
}
\address[CT]{
Istituto Nazionale Fisica Nucleare-Sezione di Catania,
Corso Italia 57, Catania 95129, Italy
}

\maketitle
\begin{abstract}
We describe some properties of nuclear ground states using a microscopic
Vlasov approach.  We propose a new method to find the energy minimum of the system
 enforcing the fermionic nature of the system.  Calculations are performed for 
two different parametrization of the mean field: a simple momentum independent
Skyrme potential and a momentum dependent one as proposed by Gale, Bertsch and Das
 Gupta.  We show that
the binding energies and radii of nuclei from $^{16}$O to $^{208}$Pb are reasonably reproduced
by opportunely tuning the surface term. We also made some calculations for neutron 
rich/poor nuclei.  
\end{abstract}


\end{frontmatter}

\section{Introduction}

Microscopic Vlasov simulations have been applied in the last two decades or so
with varying degree of success to heavy ion collisions (HIC)\cite{repo}. 
 While collective effects, particles distributions and production have been successfully
described, the approach fails when many particles correlations are important such as
in fragmentation reactions. This is of course expected since we are dealing with a
mean field approach. The method has been applied more recently to 
metallic cluster collisions and quark dynamics as well \cite{bon99}.
 One of the reasons of the success of the Vlasov approach is
surely the simplicity of its numerical solution.  Furthermore the Vlasov equation
 (VE) might include some quantum effects as well\cite{bon93}.  In that respect recall that the
one body distribution function $f({\bf r},{\bf p},t)$ entering the VE can be viewed as the
Wigner transform (WT) of the one body density matrix $\rho({\bf r},{\bf r}',t)$ which fulfills
 the Time Dependent Hartree Fock (TDHF) equation. In fact it can be easily shown
that the VE is the WT of TDHF in the limit of $\hbar \rightarrow 0$.
Thus the two approaches are practically equivalent at large entropies where quantum
effects are not important \cite{repo}. The numerical simplicity is one of the
advantages of the VE over TDHF the other is the possibility of easily including a
collision term in the VE.  Such a collision term is of course crucial for energetic
HIC.  On the other hand a serious shortcoming of the VE are the calculations of 
nuclear ground states.  While Hartree Fock (HF) is a quite well developed and accepted
method for calculations of nuclear (and other fermionic systems) ground states, there
is not so far an accepted method to obtain the ground states self-consistently in the
Vlasov approach.  This is in part due to the semiclassical nature of the VE which 
would prefer that the systems in their ground states be solids. In practical calculations
of the VE the test particles (tp) method, shortly described below, is used \cite{repo}.
Initially the tp are randomly distributed in a sphere of radius $R$ (the radius of the
nucleus) in $r$-space and of radius $P_{\rm F}$ (the Fermi momentum) in $p$-space.  Such initial
conditions are then evolved in time according to the classical equations of motion for
particles moving under the influence of a mean field.  Because of the Fermi gas initial
conditions, the system fulfills initially the requirement that the occupation factor,
i.e. the number of identical particles in a cell of size $2\pi \hbar$, must be less or
equal to one.  Such a condition is preserved in time because the VE fulfills the
Liouville theorem \cite{repo}, i.e. the density in phase space remains constant.
  While this method could be accepted in collisions of
stable nuclei at
large beam energies, it can be seriously doubted if applied to
exotic nuclei reactions and/or at beam energies close to the Coulomb barrier. It is
the purpose of this paper to show that the ground states of nuclei can be also
rather well calculated using the Vlasov approach.  This will give us the opportunity
to extend the calculations to exotic nuclei that are a quite active field now.

The paper is organized as follows.  In section 2 we review the Vlasov
 equation and the test particles method.  In section 3 we apply the
method to calculate nuclear properties with a local Skyrme potential while section 4 is devoted to
 calculations employing a momentum dependent force.  We stress the important role of the surface and
symmetry terms to reproduce some basic features of nuclei. A brief summary is given in section 5.

\section{The Vlasov Equation}

The TDHF equation of motion 
\begin{equation}
\partial_{t} \rho=[H,\rho] 
\label{tdhf}
\end{equation}
in the Wigner representation can be written as a series of the form:
\begin{equation}
\partial_{t}f+\frac{\bf p}{E}\cdot 
\nabla_{r}f-\nabla_{r}\bar U\cdot \nabla_{p}f=F_q[f] ,
\label{eq2}
\end{equation}
where $F_q[f]$ is the quantum corrections term which in the limit $\hbar \rightarrow 0$ is zero\cite{bon93},
which is the limit we will consider in this work.
The lhs is the Vlasov term and $\bar U(\bf r)$is the mean field,
which we have written in a local form but a generalization
to a momentum dependent one is straightforward\cite{repo}.

For the purpose of this work we will neglect the collision term in Eq.~(\ref{eq2})
 and note that such term will be essential when dealing with HIC.

Numerically the VE equation (\ref{eq2}) is solved by writing the one body distribution
function as:

\begin{equation}
\label{efer}
f({\bf r},{\bf p},t) = \frac{1}{n_{\rm tp}} \sum_i^N g_r({\bf r}-{\bf r}_i(t)) g_p({\bf p}-{\bf p}_i(t)) ,
\end{equation}
where the $g_r$ and $g_p$ are sharply peaked distributions (such
 as delta functions, gaussian or other simple
functions), that we shall treat as delta functions.
 $N=An_{\rm tp}$ is the number of such terms, $A$ is the mass number of the nucleus.  
Actually, $N$ is much larger than $A$, so that we can say that each nucleon
is represented by ${n_{\rm tp}}$ terms called test particles (tp).
The rigorous mean field limit
can be obtained for ${n_{\rm tp} \rightarrow \infty}$ where the calculations are
of course numerically
 impossible, even though the numerical results converge rather quickly. 
 Inserting Eq.~(\ref{efer}) in the Vlasov equation
 gives
the Hamilton equations of motions for the tp\cite{repo}:
\begin{eqnarray}
\dot {\bf r}_i &=& \frac{{\bf p}_i}{E_{i}} , \nonumber \\
\dot {\bf p}_i &=& -{{\bf\nabla}_{r_i}}U ,   \label{ripi}
\end{eqnarray}
for $ i=1,\cdots N$. The total number of tp used in this work
is more than 400000 and the results are numerically stable.

One important feature of TDHF is that the density matrix is given by a Slater determinant, i.e. 
the wave function must be properly antisymmetrized when dealing with fermions\cite{schuck}. This
implies the following property for the density matrix:
\begin{eqnarray}
\rho^2=\rho ,
\end{eqnarray}
which is preserved in time by the TDHF equation of motion. Making the Wigner transform of the
last equation in the same limit $\hbar \rightarrow 0$ gives\cite{schuck}:
\begin{eqnarray}
\bar f^2=\bar f ,
\label{eq6}
\end{eqnarray}
i.e. the occupation density $\bar f$ must be zero or one.  This is a direct consequence of the
Pauli principle which states that in a cell of size $h$ in phase space there cannot be two identical particles.
In our work we will distinguish protons and neutrons with spin up or down, thus we have four types of
particles. A consequence of this is that, for instance for $\alpha$ particles the Pauli principle does not
apply since the particles are non identical.  If one distributes randomly the tp in phase space according
to the Fermi gas model, the condition given by Eq.~(\ref{eq6}) is fulfilled and it will be preserved in time
by the VE.  But the state so obtained is not a state of minimum energy.  
Notice the analogy to the HF method in that one initially uses a Fermi gas 
for the wave functions.  
However a process of minimization is then applyed which finally gives 
the state of minimum energy\cite{schuck}.  
Such a minimization is not usually done for the VE.  
But this can be easily accomplished as follows.
First, for numerical reasons, we put a grid on the available phase space 
and store the occupation factor in each point of 
the grid with the initial conditions given by the Fermi gas model. 
We solve a frictional equation of motion with a constraint due to the 
occupation factor for each tp ($f_i$) as:
\begin{eqnarray}
\dot {\bf r}_i &=& \frac{{\bf p}_i}{E_i} +\mu_0{{\bf\nabla}_{r_i}}U ,
\nonumber \\
\dot {\bf p}_i  &=& -{{\bf\nabla}_{r_i}}U +\mu(f_i)\frac{{\bf p}_i}{E_i} ,
\label{fric}
\end{eqnarray}
where $\mu_0$ is a negative frictional coefficient independent on $f_i$,
and $\mu(f_i)$ is the $f$-dependent frictional coefficient.
$\mu(f_i)$ is positive when $f_i$ exceeds 1, while it 
is negative for $f_i<1$.
This $f$-dependence of $\mu$ is to control the local momentum space. 
The method proposed here has been first introduced in \cite{bon99} and 
refined in the Constrained Molecular Dynamics calculation of Ref.\cite{papa}.

For illustration we show in Fig.~1 the occupation factor averaged
over all the tp versus time for $^{16}$O. 
To show the power of the method we used initially a very large Fermi momentum. 
Because of that the initial occupation factor is rather small but 
it increases quickly and stabilizes around 1 after about 200 fm/$c$.  
Results for different nuclei and forces are quite similar to the
one illustrated in Fig.~1. 

\begin{figure}[tbp]
\epsfysize=7 truecm \epsfbox{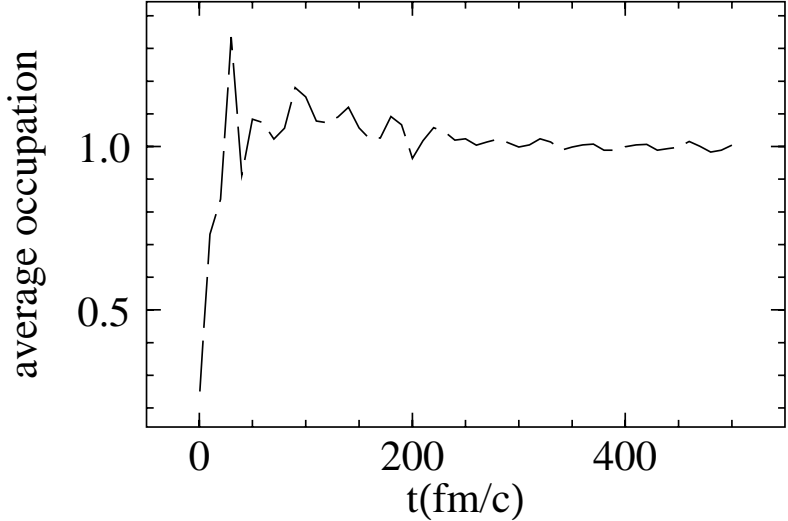}
\caption{Average occupation in phase space versus time for $^{16}$O.\label{fig:fig1}}
\end{figure}

\section{Local Mean Field}

In addition to the Coulomb interaction for protons, a popularly used parametrization of the mean field
is given by the Skyrme parametrization:
\begin{eqnarray}
\bar U(\rho({\bf r}))=a\rho({\bf r})+b\rho^\gamma({\bf r})+c_s\nabla^2\rho+c_a\frac{(\rho_n-\rho_p)^2}{2\rho_0} ,
\end{eqnarray}
where $\rho$,$\rho_{n,p}$ are the total, neutron and proton density respectively,  $\rho_0=0.16\ {\rm fm}^{-3}$. 
The parameters a,b and n are tuned to have a compressibility of 225 MeV\cite{repo}. The canonical values
of the parameters entering the surface and symmetry terms are $c_s=-138\ {\rm MeV\,fm}^5$ and $c_a=32$ MeV respectively.
However, when using a grid in phase space a surface term is spuriously introduced.  Because of that we fixed
the value of $c_s=6.9\ {\rm MeV\, fm}^5$ by fitting the binding energy of $^{16}$O, i.e. a small nucleus for which the surface
is very important while the symmetry energy is negligible.  The small value of the surface term coefficient should not
surprise, in fact for density calculations in $r$-space we use a grid of 1 fm side.  This introduces a surface term
 spuriously.  Changing method of solution, for instance gaussians instead of delta function, will 
give rise to a surface term from the volume part of the mean field.  Also in that case one must adjust the
surface term to some known property of finite nuclei as we did here.

In Fig.~2 we plot the binding energies versus time for four different systems.  The dashed line 
indicates the value of the binding energy obtained from the Weizsacker formula \cite{schuck} whose
parameters are fitted to stable nuclei.  We stress that all the terms entering the Weizsacker formula
 are in principle contained in out approach excluding the pairing energy.  We see that the BE from systematics
are reasonably
 reproduced by the method.

In Fig.~3 we plot the radius versus time for the systems as above.  The full lines refer to neutrons and
the dashed lines to protons radii respectively.  A neutron halo is quite evident for the unstable C, while the
proton radius is slightly larger than the neutron one for the Sn nucleus.  The Pb nucleus gives  similar radii
which is due to the competing action of the symmetry and coulomb terms.  It also seems that the neutron and
 protons oscillate with different frequencies.  This is especially seen in the C case, which indicates that the
protons and neutrons have different compressibilities.  We would like to notice also that in the Pb case, due
to the strong Coulomb force it is quite difficult to reach a state of minimum energy.  Because of the relative
small number of tp per nucleon (about 4000), classical fluctuations could lead the system to fission if the
minimization procedure is followed for a longer time.

\begin{figure}[tbp]
\epsfysize=9 truecm \epsfbox{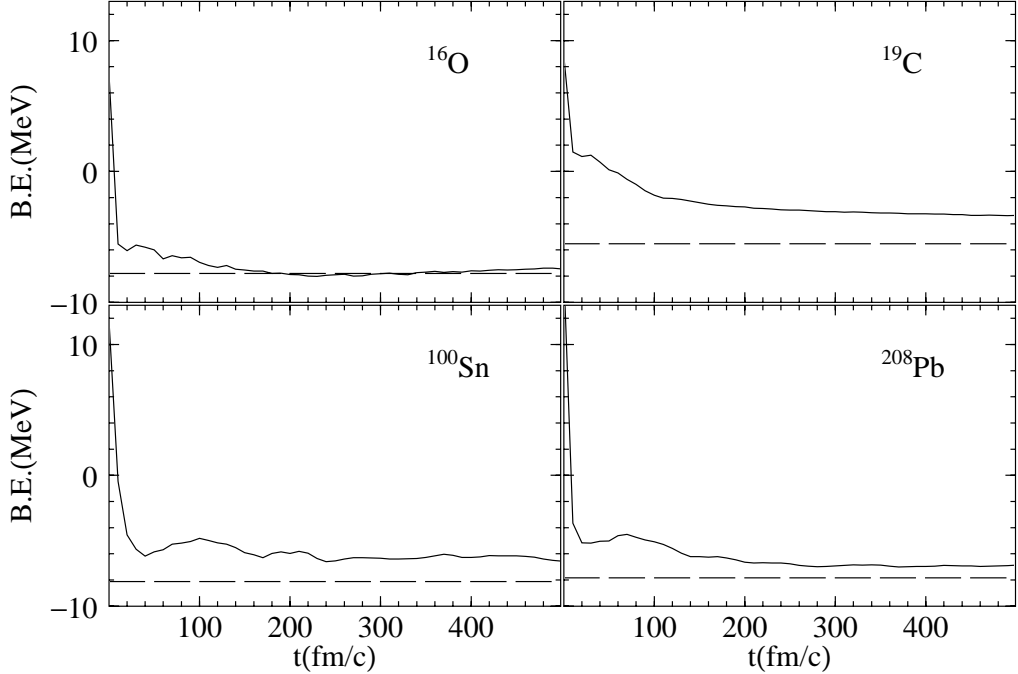}
\caption{Binding energy per nucleon versus time for different nuclei as indicated. The dashed line
is the Weizsacker parametrization fitted to stable nuclei. \label{fig:fig2}}
\end{figure}

\begin{figure}[tbp]
\epsfysize=9 truecm \epsfbox{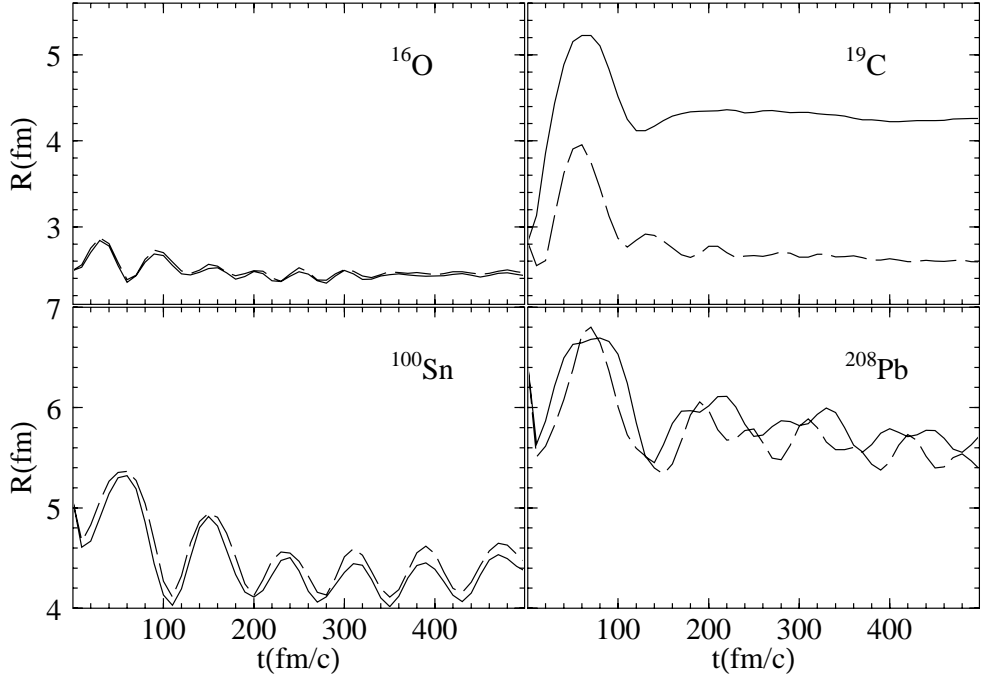}
\caption{Proton (dashed lines) and neutron (full lines) radii versus time. \label{fig:fig3}}
\end{figure}

\section{Momentum Dependent Interaction}

It is well known that the nuclear interaction is momentum dependent.  Many expression have been 
proposed, here we use one introduced by C. Gale et al.\cite{gale}:
\begin{eqnarray}
\bar U(\rho({\bf r}),{\bf p})=\bar U(\rho)+\frac{2c}{\rho_0}\int{d^3p'\frac{f({\bf r},{\bf p}')}{1+[|{\bf p}-{\bf p}'|/\Lambda]}} .
\end{eqnarray}
The values of the parameters are given in \cite{gale}.  Notice, that also in this case the surface term
was adjusted to reproduce the binding energy of $^{16}$O. It turns out that the value of $c_s$ is half of
the canonical one.

In Figs.~4 and 5 we plot the BE and radii of the nuclei as in Figs.~2 and 3.  Also in this case convergency
is rapidly obtained and the BE are in good agreement with the Weizsacker parametrization \cite{schuck}
(dashed lines). The behavior of the radii is as before with a neutron halo appearing in the  $^{19}$C
 case.  Also notice the different oscillations for neutron and protons for this exotic nucleus.

\begin{figure}[tbp]
\epsfysize=9 truecm \epsfbox{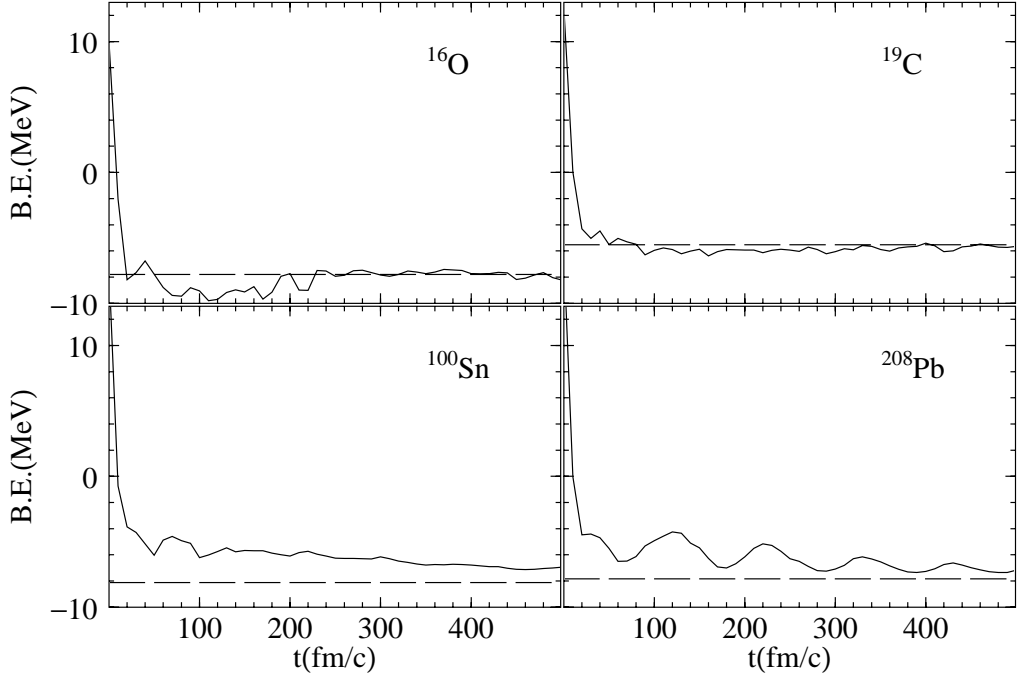}
\caption{Same as in figure 2 but for the momentum dependent interaction.
\label{fig:fig4}}
\end{figure}

\begin{figure}[tbp]
\epsfysize=9 truecm \epsfbox{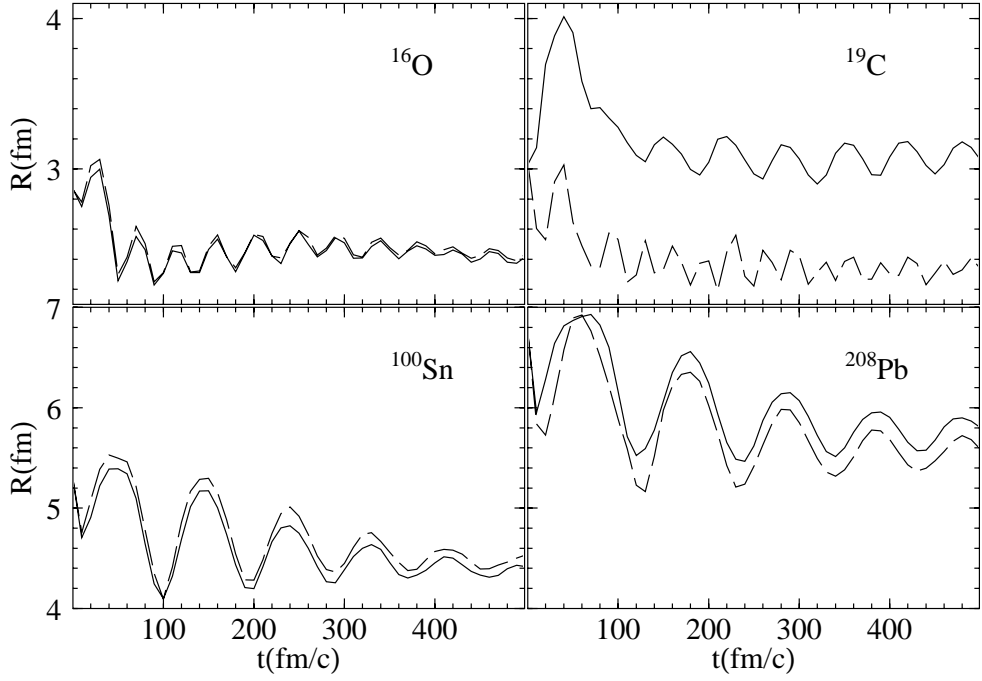}
\caption{Same as in figure 3 but for the momentum dependent interaction.
\label{fig:fig5}}
\end{figure}

\section{Conclusions}

In this paper we have presented a novel approach to get fermion ground states within a Vlasov
 approach.  This approach is the semiclassical counterpart of the HF method.  It has some advantages 
compared to HF in a simpler numerical solution and a much easier extension to collision problems.  On the
other hand some quantum effects are not included.  This is especially important when the mean field is
not a smooth function.  As in all mean field calculations symmetries are preserved and because of that,
 with the use of the mean field parametrization of this paper, nuclei are spherical symmetrical.  Correlations
might break 
symmetries but these are not contained here.  In collision problems one can easily include 
a two body collision term.  The approach might be suited for calculating low energy phenomena such
as fusion, deep inelastic, particle production and collective flow at energies up to few hundreds MeV/nucleon.


\begin{thebibliography}{99}

\bibitem{repo} A.~Bonasera, F.~Gulminelli and J.~Molitoris, 
Phys. Rep. {\bf 243}, (1994)1;  G.~Bertsch, S.~Das Gupta, 
Phys. Rep. {\bf 160} (1988) {189}, and references therein.

\bibitem{bon99} A.~Bonasera, nucl-th/9905025; nucl-th/9908036 and 
Phys. Rev. {\bf C60} (1999) 065212;
Phys. Rev. {\bf C62} (2000) 052202(R); Nucl. Phys. {\bf A681} (2001) 64c.

\bibitem{bon93} A.~Bonasera, V.N.~Kondratyev, A.~Smerzi and E.A.~Remler,
Phys. Rev. Lett. {\bf 71} (1993) 505.
  
\bibitem{schuck}
P.Ring, P.Schuck, {\it The Nuclear Many-Body Problem},Springer-Verlag,
New York, 1980.

\bibitem{papa}M.~Papa, T.~Maruyama and A.~Bonasera, 
Phys. Rev. C {\bf C64} (2001) 024612.


\bibitem{gale} C.~Gale, G.~Bertsch and S.~Das Gupta, Phys. Rev. {\bf C35} (1987) 1666.


\end{thebibliography}
\end{document}